\documentclass[aps,prl,twocolumn,showpacs,a4paper]{revtex4}
\usepackage{graphicx}

\begin{document}





\title{Effect of the shot-noise on a Coulomb blockaded single
Josephson junction}
\author{E.B. Sonin$^{1,2}$}

\affiliation{$^{1}$Racah Institute of Physics, Hebrew University of
Jerusalem, Jerusalem 91904, Israel  \\
$^{2}$Low Temperature Laboratory, Helsinki University of Technology, FIN-02015 HUT, Finland }

\date{\today}

\begin{abstract}

We have investigated how the Coulomb blockade of a mesoscopic Josephson junction in
a high-impedance environment is suppressed by shot noise from  an
adjacent junction. The presented theoretical analysis is an extension of the phase
correlation theory for the case of a non-Gaussian noise. Asymmetry of the
non-Gaussian noise should result in the shift of the conductance minimum from zero
voltage and the ratchet effect (nonzero
current at zero voltage), which have been experimentally observed. The analysis
demonstrates that a Coulomb blockaded tunnel junction in a high impedance
environment can be used as an effective noise detector.
\end{abstract}



%



\maketitle

The Coulomb blockade is very sensitive to fluctuations in an environment. The
Johnson-Nyquist noise results in a power-law-like increase
of conductance as a function of temperature \cite{Aver,SZ,IN}: $G \propto T^{2\rho
-2}$. The exponent of the power law, $2\rho-2$, is governed by the parameter
$\rho=R/R_Q$ where the resistance $R$ describes the dissipative ohmic environment
and $R_Q=h/4e^2$ is the quantum resistance. Tuning the dissipative parameter $\rho$
one expects the quantum or dissipative transition (superconductor--insulator
transition) at $\rho=1$, which was revealed experimentally \cite{Penttila}.
Then in the case of large exponents $2\rho-2 \gg 1$ (the insulator state), there
is a high resolution against tiny changes in temperature, or alternatively, a high
sensitivity to thermal noise.

One would expect that the Coulomb blockaded tunnel junction would be sensitive
to other noise sources as well. In Ref. \onlinecite{SN} the Coulomb blockade of
Cooper pairs was used to detect shot noise induced by a
separately biased superconductor-insulator-normal metal (SIN) tunnel
junction. The current through SIN was
found to strongly reduce the Coulomb blockade of Cooper pairs. This  has shown a way
to use a Josephson tunnel junction for ``noise spectroscopy'', and other modifications of
this method of noise investigation have been discussed in the literature \cite{SNO}.
However, the theoretical analysis given in Ref.
\onlinecite{SN} was far from complete: (i)
While the experiment was done in the insulator state $\rho \gg 1$, the theoretical
analysis was performed for $\rho <1$ using
the perturbation theory with respect to shot noise.(ii) The theory could not
explain the asymmetry of observed $IV$ curves 
(shift of the conductance minimum from zero voltage).

The present Letter suggests a theory for the effect of shot noise from an
independent source on a Coulomb blockaded Josephson junction. The junction is in
the insulator state with high impedance environment $\rho \gg 1$, which is most
suitable for purposes of noise detection. The theory shows 
that since shot noise is non-Gaussian and asymmetric \cite{Les}, the $IV$ curves should also
be asymmetric, i.e., conductance is not an even function of voltage. This should result
in the shift of the conductance minimum and the ratchet effect (a finite current at
zero voltage bias on the junction \cite{BG}), which have already been observed
experimentally \cite{exp}. Asymmetry of
$IV$ curves originates from odd moments of the shot noise, which are now
intensively discussed theoretically \cite{Odd}, but are quite difficult to detect
experimentally by other methods \cite{OddE}. This is a manifestation
of rich possibilities of noise detection with Coulomb
blockaded  junctions. The present theory addresses shot noise in a
Josephson junction, but apparently it can be generalized to other types of
noise and to a Coulomb blockaded normal junction. So the latter can be used for 
noise detection as well.

\begin{figure}
  \begin{center}
    \leavevmode
    \includegraphics[width=0.9\linewidth]{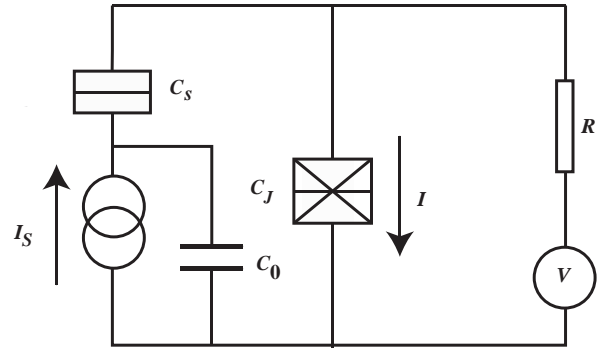}
    \caption{Electric circuit. }
  \label{fig1}
  \end{center}
  \end{figure}

The electric circuit for our analysis is shown in Fig. \ref{fig1}. The
Josephson tunnel junction  with capacitance $C_J$ is voltage-biased via the shunt
resistor $R$. Parallel to the Josephson junction there is another junction
(noise junction) with the capacitance $C_s$ connected with an independent
DC current source. The role of very large capacitance $C_0$ is to shortcircuit the large ohmic resistance of
the current source for finite frequencies. The current
$I_S$ through the noise junction produces shot noise, which affects the $IV$
curve of the Josephson junction. Let us recall well known results concerning the
$IV$ curve without shot noise ($I_S=0$). We assume that the  Josephson coupling
energy $E_J\cos \varphi$ ($\varphi$ is the Cooper-pair phase across the Josephson junction)
is weak in comparison with the Coulomb energy $E_C =e^2/2C$, where
$C=C_s+C_J$, and the former can be considered as a
perturbation. Then the
Golden Rule gives for the current  of Cooper pairs \cite{Aver,IN}:
\begin{equation}
I={\pi eE_J^2 \over \hbar}[P(2eV)-P(-2eV)]~, 
 \end{equation}
where the function
\begin{equation}
P(E)={1\over 2\pi \hbar}\int_{-\infty}^\infty dt\, e^{iEt}\left\langle
e^{i\varphi(t_0)}e^{-i\varphi (t_0-t)}\right\rangle~ 
 \end{equation}
characterizes the probability to excite an environment mode with the energy $E$. Since we use the
perturbation theory with respect to $E_J$ we can calculate phase fluctuations neglecting $E_J$, i.e.,
treating the Josephson junction as a capacitor. The crucial assumption in the phase-fluctuation theory (or
$P(E)$ theory) is that the phase fluctuations are Gaussian \cite{IN} and 
\begin{equation}
\langle e^{i\varphi(t_0)}e^{-i\varphi (t_0-t)}\rangle \approx e^{J_0(t)} ~,
  \label{Gauss}    \end{equation}
where the phase--phase correlator
\begin{eqnarray}
J_0(t) =\langle[ \varphi(t)
-\varphi(0)]\varphi(0)\rangle=J_R(t)+ iJ_I(t) \nonumber \\
= 2\int_{-\infty}^\infty {d\omega \over \omega}
\frac{\mbox{Re} Z(\omega)}{R_Q}
 \frac{e^{-i\omega t} -1}{1-e^{-\beta \hbar \omega}}
      \end{eqnarray}
is a complex function of time determined by the Johnson-Nyquist
noise in the environment, i.e., in the electric circuit with the impedance
$Z(\omega)$. Here $\beta =1/k_BT$ is the inverse temperature. Then the current
is
\begin{eqnarray}
I=-{2 eE_J^2 \over \hbar^2}\mbox{Im}\int_0^\infty dt e^{J_0(t)}
 \sin \left(2eVt\over \hbar\right)
 \label{WC}  \end{eqnarray}
and the zero-bias conductance is given by
\begin{equation}
G_0=\left.{dI\over dV}\right|_{V\rightarrow 0}=-{4 e^2E_J^2 \over
\hbar^3}\mbox{Im}\int_0^\infty t\,dt e^{J_0(t)} \,.
          \label{G-0}\end{equation}
In our case  $Z(\omega) =(1/R +i\omega C)^{-1}$, and at $T \to 0$
($t>0$):
\begin{eqnarray}
J_0(t) =  \rho \left[-e^{t/\tau}\mbox{E}_1\left({t\over \tau}\right)-
e^{-t/\tau}\mbox{E}_1\left(-{t\over \tau}+i0\right) \right.\nonumber \\ \left.
-2 \ln{ t \over \tau}
 -2\gamma -i\pi\right] \approx  \rho \left[{t^2  \over \tau^2}\left(\ln{t\over
\tau}+\gamma -{3\over 2}\right) -{i\pi t\over \tau}\right]\,,
     \label{J-0}     \end{eqnarray}
 where $\tau=RC$, $\gamma=0.577$ is the Euler constant, and
$\mbox{E}_1(z)=\int_1^\infty (e^{-zt}/t)\,dt$ is the exponential integral
\cite{AS}. The second expression yields the short-time expansion ($t\ll \tau$).
The small imaginary correction $+i0$ to the argument of one of the
exponential integrals is important for analytic continuation of $\mbox{E}_1(z)$ from real
$t$ to the complex plane \cite{AS}. At long times $t \gg\hbar
\beta\gg\tau$ one cannot ignore the temperature, which determines the phase diffusion:
$J_R(t)
\approx - 2\pi\rho  t /\hbar\beta=- 2\pi\rho k_BT t /\hbar$. In the insulator
state $\rho >1$ in the zero-temperature limit the conductance
$G(T) \propto T^{2(\rho-1)}$ vanishes, and the current depends nonanalytically on
voltage: $I \propto V^{2\rho-1}$. 

One can take into account shot noise  simply by adding the 
shot noise contribution to the correlator $J_0(t)$.
Then the fluctuating phase $\varphi=\varphi_0+\varphi_s$ consists of
two terms: $\varphi_0$  from Johnson-Nyquist noise, and
$\varphi_s$ from shot noise. Since the fluctuations are
uncorrelated the phase--phase correlator is a superposition of two noise
contributions: $J(t)=J_0(t)+J_s(t)$.   At long times $t
\gg \tau$ the shot noise  modifies the phase diffusion: $J_0(t)+J_s(t)
=-2\pi\rho k_B( T +T_N)t /\hbar$, where $T_N =e|I_S |R/2k_B$ is the
noise temperature. As a result, the conductance power-law expression changes: $G \sim
(T+ T_N)^{2\rho -2}$. At $T=0$ this yields a nonanalytic
power-law dependence on the noise current: $G\propto |I_S|^{2\rho-2}$.
But this approach used in Ref.
\cite{SN} fails at $\rho \gg 1$, which is the most interesting case for noise
detection. It ignores the odd moments of shot noise, since the
correlator
$J_s(t) =\langle[ \varphi_s(t)
-\varphi_s(0)]\varphi_s(0)\rangle$ is quadratic with respect to $\varphi_s$.

In the theory for the case $\rho \gg 1$ one should abandon Eq.
(\ref{Gauss}), which assumes that the noise is Gaussian. On the
other hand, we assume that the fluctuation
$\varphi_s$ is classical and the values of  $\varphi_s$ at different moments of
time commute. Since the equilibrium noise and the shot noise are uncorrelated,  the
generalization of Eq. (\ref{WC}) is 
\begin{eqnarray}
I=-{2 eE_J^2 \over \hbar^2}\mbox{Im}\left\{\int_0^\infty dt e^{J_0(t)}
\left\langle \sin \left({2eVt\over \hbar}+\Delta
\varphi_s\right)\right\rangle\right\} .
 \label{curG} \end{eqnarray}
Here $\Delta \varphi_s= \varphi_s(t_0)-\varphi_s(t_0-t)$. Subtracting from (\ref{curG}) the current at
$\Delta \varphi_s=0$ given by Eq. (\ref{WC}) we receive the shot-noise contribution to the current (the only
contribution at 
$T
\to 0$):
\begin{eqnarray}
\Delta_sI =-{2 eE_J^2 \over \hbar^2}\mbox{Im}\left\{\int_0^\infty dt e^{J_0(t) }
\left[\cos {2e V t\over \hbar}\left\langle \sin \Delta
\varphi_s\right\rangle
\right. \right.\nonumber \\ \left. \left.
+\sin {2e V t\over \hbar}(\left\langle \cos \Delta \varphi_s\right\rangle
-1)\right]\right\}\,. 
        \label{cur}       \end{eqnarray}

Now we want to calculate the shot-noise fluctuations $\varphi_s$.
The charge transport through the noise junction is a sequence of current
peaks $\delta I=\mbox{sign}(I_S)e\sum_i\delta (t-t_i)$, where $t_i$ are the random
moments of time when an electron crosses the junction \cite{Blanter}. We neglect the
duration of the tunneling event itself. The positive sign of
$I_S$ corresponds to the current shown in Fig. \ref{fig1}.  Any peak
generates a voltage pulse  at the Josephson junction:
$V_s(t)=\mbox{sign}(I_S)(e/C)\sum_i \Theta (t-t_i) e^{-(t-t_i)/\tau}$ where
$\Theta(t)
$ is the step function. After using the Josephson relation $\hbar \partial \varphi
/\partial t=2eV$ one finds that $\varphi_s$ is given by a sequence of phase jumps at the Josephson
junction: 
\begin{eqnarray}
\varphi_s(t)  =\mbox{sign}(I_S){2e^2\tau \over \hbar C}\sum_i \Theta (t-t_i)\left [
1- e^{-(t-t_i)/\tau}\right] \nonumber \\
=\mbox{sign}(I_S)\pi\rho \sum_i \Theta (t-t_i)\left [ 1-
e^{-(t-t_i)/\tau}\right]  ~.
     \end{eqnarray}
The phase difference between two moments $t_0$ and $t_0-t$ can be presented as 
$\Delta \varphi_s=\sum _i 
\delta \varphi_s(t,t_0-t_i)$, where
\begin{eqnarray}
\delta \varphi_s(t,\tilde t) 
=\mbox{sign}(I_S)\pi \rho\left\{ \Theta (\tilde t)\left [ 1-
e^{-(\tilde t)/\tau}\right]\right.
\nonumber \\ \left.
- \Theta (\tilde t-t)\left [ 1-
e^{-(\tilde t-t)/\tau}\right]\right\} ~.
     \end{eqnarray}
is the phase difference generated by a single current peak and $\tilde t=t_0-t_i$. Since 
phase jumps are not  small for
$\rho \gg 1$, one cannot expand the sine
and cosine functions in Eq. (\ref{cur}).

For small currents $|I_S| \ll e/\tau $ through the noise junction the voltage pulses
and phase jumps are well separated in time. Then the average powers of the phase
difference can be approximately estimated as 
\begin{eqnarray}
\left\langle \Delta \varphi_s ^n
\right \rangle=\sum _i  \left\langle[\delta \varphi_s(t,t_0-t_i)]^n \right
\rangle  \nonumber \\
= {|I_S| \over e}\int_{-\infty}^\infty d \tilde t
 [\delta \varphi_s( t, \tilde t)]^n~.
  \label{PL} \end{eqnarray}
The averaging is performed over the random sequences of moments $t_i$, which
eliminates any dependence on the time $t_0$. In Eq. (\ref{PL}) we neglected cross-terms,
which contain the phases from more than one pulse and are proportional to
higher powers of a small parameter
$|I_S|\tau /e$. 

Equation (\ref{PL}) can be extended on any function of $\Delta \varphi_s$, which
vanishes at $\Delta \varphi_s \rightarrow 0$,
and hence we obtain
\begin{eqnarray}
\langle \sin \Delta \varphi_s \rangle = {|I_S| \over e}\int_{-\infty}^\infty d\tilde
t
\sin \delta \varphi_s(t,\tilde t)
\nonumber \\ 
= {I_S\tau \over e} \left\{{\pi \over
2}+ \mbox{si} \left[r\left(1-
e^{-t/\tau} 
\right) \right]  \right.\nonumber \\ \left.
+\sin r
\left[\mbox{ci} r -
\mbox{ci} \left(r  e^{-t/\tau} \right)\right] -\cos r
\left[\mbox{si} r - \mbox{si} \left(r 
e^{-t/\tau} \right)\right] \right\},
              \end{eqnarray}
\begin{eqnarray}
\langle \cos \Delta \varphi_s \rangle-1={|I_S| \over e}\int_{-\infty}^\infty d\tilde
t
\left[\cos \delta \varphi_s(t, \tilde t) -1\right]\nonumber \\
= {|I_S|\tau \over e} \left\{-{t \over
\tau} 
+\cos r
\left[\mbox{ci} r -
\mbox{ci} \left(r  e^{-t/\tau} \right)\right]  \right.\nonumber \\ \left.
+ \sin r
\left[\mbox{si} r - \mbox{si} \left(r 
e^{-t/\tau} \right)\right]    
+\mbox{ci} \left[r\left(1-
e^{-t/\tau} 
\right) \right]  \right.\nonumber \\ \left.
-\gamma-\ln \left[r\left(1-
e^{-t/\tau} 
\right) \right] \right\}~, 
     \label{cos}         \end{eqnarray}
where $\mbox{si}(x)=-\int_x^\infty
\sin t\, dt/t$ and $\mbox{ci}(x)=-\int_x^\infty \cos t\, dt/t$ are sine and cosine
integral functions, and $r=\pi \rho$. 

Analyzing the $IV$ curve at small voltage bias, one can expand
Eq. (\ref{cur} ) in  $V$:
\begin{equation}
\Delta_sI=I_0 +G_s V+ a V^2+bV^3~,
      \end{equation}
where 
\begin{eqnarray}
G_s=-{ 4e^2 E_J^2 \over \hbar^3 }\mbox{Im} \int_0^\infty t\, dt e^{J_0(t)
}[\left\langle\cos \Delta\varphi_s\right\rangle-1]
               \end{eqnarray}
is the shot-noise conductance, the constant current
\begin{eqnarray}
I_0=-{ 2e E_J^2 \over \hbar^2 }\mbox{Im} \int_0^\infty  dt e^{J_0(t)
}\left\langle\sin \Delta\varphi_s \right\rangle
               \end{eqnarray}
determines the ratchet effect, and the constants 
\begin{eqnarray}
a={ 4e^3 E_J^2 \over \hbar^4}\mbox{Im} \int_0^\infty t^2\, dt e^{J_0(t)
}\left\langle\sin \Delta\varphi_s \right\rangle
               \end{eqnarray}
and
\begin{eqnarray}
b={ 8 e^4 E_J^2 \over 3 \hbar^5}\mbox{Im} \int_0^\infty t^3\, dt e^{J_0(t)
}[\left\langle\cos \Delta\varphi_s\right\rangle-1]
               \end{eqnarray}
determine the curvature of the  conductance-voltage plot and the shift of the
conductance minimum.

In the high-impedance limit $\rho \gg 1$ it is possible to calculate the parameters
of the $IV$ curve analytically. As one can see below, the most important
contribution to the integrals comes from $t \sim \tau /\sqrt{r}$. Since
$r=\pi \rho$ is large, these values of $t$ are small compared to $\tau$ and one
can use the small-argument expansion for the Johnson-Nyquist correlator given in Eq.
(\ref{J-0}). On the other hand the arguments $r t/\tau \sim  \sqrt{r}$ of the sine and the
cosine integrals are large and one should use asymptotic expansions for them:
$\mbox{si}(x) \sim -\cos x/x $,
$\mbox{ci}(x) \sim \sin x/x
$. Let us consider, for example, the integral for $G_s$:
\begin{eqnarray}
G_s\approx -{ 4e E_J^2 \tau \over \hbar^3 }|I_S|\int_0^\infty t\, dt \mbox{Im}
\left\{e^{J_0(t)}\right\}{\sin 
\left[r\left(1-e^{-t/\tau} \right) \right] \over r\left(1- e^{-t/\tau} \right)} \nonumber
\\ 
\approx { 4e E_J^2 \tau^2\over \hbar^3 r}|I_S| \int_0^\infty  dt
\exp \left(-\rho{t^2  \over \tau^2}\ln{\tau \over
t}\right) \sin ^2 {rt \over\tau}  
\nonumber \\ 
\approx  {\sqrt{2} \pi e E_J^2 \tau^3\over \hbar^3 r^{3/2}\sqrt{\ln r}}|I_S|
\approx {\pi^{5/2}E_J^2 C^3 \over 4
e^5 \sqrt{2\ln \rho}}\rho^{3/2}|I_S|~.
    \label{integr}           \end{eqnarray}
In a similar way one can calculate the integrals determining the other parameters
of the IV-curve:
\begin{eqnarray}
I_0= {\pi^2 E_J^2 C^2\over 8e^4 }\rho I_S,~a={I_S \over |I_S|}{  C\over 2e }G_s,~b=-{\pi^2 C^2\rho\over 6
e^2\ln \rho}G_s.
          \end{eqnarray}

If the current $I_S$ through the second (noise) junction is not small compared to
$e/\tau$ the voltage (phase) pulses start to overlap. Eventually at further growth
of the current, so that $|I_S| \gg e/\tau$, the current fluctuations around the
average value become very small. In this limit with a good accuracy the phase shift
$\Delta \varphi_s (t) $ is a linear function of time, proportional to the voltage
drop $I_S R$ on the shunt resistance, and the voltage at the Josephson junction is
$ V+I_S R$. Then as a function of the voltage bias $V$ the $IV$ curve would be
asymmetric. In particular, the conductance would have  a minimum at  $V_{min}=
-I_S R$, and not at
$V=0$.  However, for small
$I_S$ when the electron transport through the noise junction is a sequence of well
separated current pulses the asymmetry parameters are essentially different from
the electric-circuit effects at large currents, even though they have the same
signs. 

The ratchet effect can be quantified by the ratio of currents (the current
gain) at zero voltage bias: $ \beta_0=I_0/ I_S$.
Our analysis is valid until $\beta_0 <1$. If $\beta_0
\gg 1$, a single current pulse at the noise junction is able to produce many Cooper
pair tunnelings through the Josephson junction. But then the shot noise at the
latter junction would be more important than the noise at the former. Moreover,
according to Ref. \onlinecite{IN} (Sec. 5.2) the condition $\beta_0 < 1$ is required  for
validity of the perturbation theory used to calculate the current through the
Josephson junction.

In contrast to the equilibrium Johnson-Nyquist-noise governed conductance, which  is a nonanalytic
function of $V$  in the $T=0$ limit determined by long-time correlations \cite{Aver,SZ,IN}, the shot-noise
governed conductance  for large $\rho$ has a leading analytic contribution, which comes
from short times $\sim
\tau/\sqrt{\rho}$ as evident from Eq. (\ref{integr}). Without shot noise the analytic contribution exactly
vanishes. One can see it by rotating the integration path in Eq. (\ref{G-0}) from the real to the
negative imaginary axis  in the complex plane. After rotation the integral becomes purely real while the
conductance is determined by the imaginary part of the integral. For the purely equilibrium noise the
approximation used in Eq. (\ref{integr}) is invalid since  the integrand is a strongly oscillating
function. But even in the presence of shot noise one can use Eq. (\ref{integr}) only until the conductance
integral is convergent at long times $t\gg
\tau$. In this limit according to Eq. (\ref{J-0} ) the Johnson--Nyquist correlator
$J_0(t) \propto -2\rho \ln t$ and the shot noise correlator $\left\langle
\cos\Delta \varphi_s\right \rangle \propto t$ [see Eq. (\ref{cos})]. Then the
integral for the conductance $G_s$  is convergent only if $\rho >1.5$. As long as this inequality is
satisfied, $G_s$ is linear with respect to $|I_S|$. But for $\rho < 1.5$ the nonanalytic contribution 
 $G \propto |I_S| ^{2\rho -2}$, which was derived in Ref. \onlinecite{SN}, becomes more important. Thus the
transition from nonanalytic to analytic behavior of $G_s$ should occur at  $\rho=1.5$.

We assumed that the current $I_S$ was small, but not
so small that Johnson-Nyquist noise from the noise junction could compete with the
shot noise. This is achieved by a high resistance $R_s$ of the noise junction:
$R_sI_S\gg k_BT/e~,~ \hbar \omega/e$ \cite{PH}. Here the relevant
frequency is $\omega  \sim \sqrt{\rho}/\tau$. These conditions are satisfied
in the experiment, which is compared with
our theory  \cite{exp}.

In summary, we have presented a theory of
the  effect of shot noise from an independent source on the Coulomb blockaded
Josephson junction in a high-impedance environment. Though we used the framework of
phase correlation theory, we essentially modified it by admitting that
fluctuations are not Gaussian. The analysis takes into account the asymmetry of 
shot noise characterized by its odd moments. For
high impedance environment the effect is so strong that the expansion in moments is not valid and was not
used in the analysis. The asymmetry of the shot noise results in asymmetry of the $IV$ curve:
the shift of the conductance minimum from the zero bias and the ratchet effect, which
have been observed experimentally \cite{exp}. Altogether, the analysis demonstrates, that the
Coulomb blockaded tunnel junction in a high impedance environment  can be employed as a sensitive
detector of shot noise. We expect that other sources of noise (e.g.,
$1/f$--noise), produce similar effects. 

I acknowledge collaboration with Julien Delahaye, Pertti Hakonen,
Tero Heikkil\"a, Rene Lindell, Mikko Paalanen, and Mikko Sillanp\"a\"a. Also I
appreciate interesting discussions with Dmitry Averin, Marcus B\"uttiker, Daniel Esteve,
Joseph Imry, Yeshua Levinson, Yuval Oreg, Michael Reznikov, Cristian Urbina, Andrei Zaikin, and Alexander
Zorin. The work was supported by the Academy of Finland, by the Large Scale Installation Program ULTI-3 of
the European Union, and by the grant of the Israel Academy of Sciences and Humanities.

\end{document}